\documentstyle[10pt,conf_iap,epsf]{article}

\def\kms{{\rm\, km\ s^{-1}}}

\def\mpc{{\rm\,Mpc}}
\def\msun{{\rm\,M_\odot}}
\def\eg{{\it e.g., }}
\def\ie{{\it i.e., }}
\def\etal{{\it et al. }}
\def\etc{{\it etc. }}

\def\spose#1{\hbox to 0pt{#1\hss}}
\def\lta{\mathrel{\spose{\lower 3pt\hbox{$\mathchar"218$}}
     \raise 2.0pt\hbox{$\mathchar"13C$}}}
\def\gta{\mathrel{\spose{\lower 3pt\hbox{$\mathchar"218$}}
     \raise 2.0pt\hbox{$\mathchar"13E$}}}

\begin{document}

\heading{%
%
Lyman Alpha Absorption in The Cosmic Web
} 
\par\medskip\noindent

\author{J. Richard Bond and James W. Wadsley}
\address{%
Canadian Institute for Theoretical Astrophysics and 
Department of Astronomy, \\
University of Toronto, 60 St. George St., Toronto,
ON M5S 3H8, Canada}

\begin{abstract}
We describe large scale structure at high redshift in terms of the
Cosmic Web picture for $\{S,\Lambda,O,H\}$CDM models: how
galactic-scale ``peak-patches'', filaments and membranes create an
interconnected intergalactic medium. The ideas are applied to our
Ly$\alpha$ forest simulations of ``shear-field patches''. We discuss
simulation method and design, resolution dependence, the statistical
combination of patches, UV flux scaling, and whether filtered
Zel'dovich maps are useful. The response to changes in power spectrum
shape and amplitude, and in cosmological parameters, is described. We
also show $\Omega_b {\rm h}^2$ derived from UV rescaling is
overestimated if the resolution is not adequate.
\end{abstract}

\section{The Cosmic Web and Peak-Patches at High Redshift} 

The Cosmic Web picture~\cite{bkp96} provides a powerful language for
understanding the structure and evolution of Lyman Alpha absorption
systems~\cite{bwKing97}. It predicts the basic structural components
of the IGM as a function of scale, epoch and cosmology. For density
contrasts $\delta \gsim 100$, the rare-events at $z\sim 3$ are massive
galaxies, observed as damped Ly$\alpha$ absorbers, while the typical
collapsed objects are dwarf galaxies responsible for Lyman Limit and
metal line systems. The medium is visually dominated by $\delta \sim
5-10$ filaments, bridging massive galaxy peaks, with smaller scale
filaments within the larger scale ones bridging smaller dwarf galaxies
contributing the most to the $N_{HI} \lta 10^{14.5}$ Ly$\alpha$
forest.


$N$-body calculations cannot currently cover large enough volumes of
space with sufficient resolution to simulate high redshift galaxy
catalogues (with clustering) to compare with \eg the Steidel \etal
structure at $z \sim 3.1$ \cite{steidel97}. Only intermediate and
small scale structures are accessible to direct gasdynamical
simulations of the Ly$\alpha$ forest. A peak-patch catalogue with
ultralong waves included can treat such large volumes quickly and
accurately~\cite{bm96}. Galaxy halos are identified with regions of
space determined to have collapsed using ellipsoidal internal
dynamics, with external tidal fields playing a large role. Bulk
properties like mass, binding energy, velocity and shear for the
patches follow, and can be used with single-patch hydrodynamics or
phenomenology to predict internal gas profiles. Field realizations
constrained to have interesting multiple peak/void structures can be
used as initial conditions for high resolution numerical simulations,
{\it e.g.} the strong filament of galaxies in~\cite{bwKing97}.

\begin{figure}[p]
\vspace{-1.5in}
\centerline{
\hspace{1.4in}
\epsfysize=3.5in\epsfxsize=3.5in\epsfbox{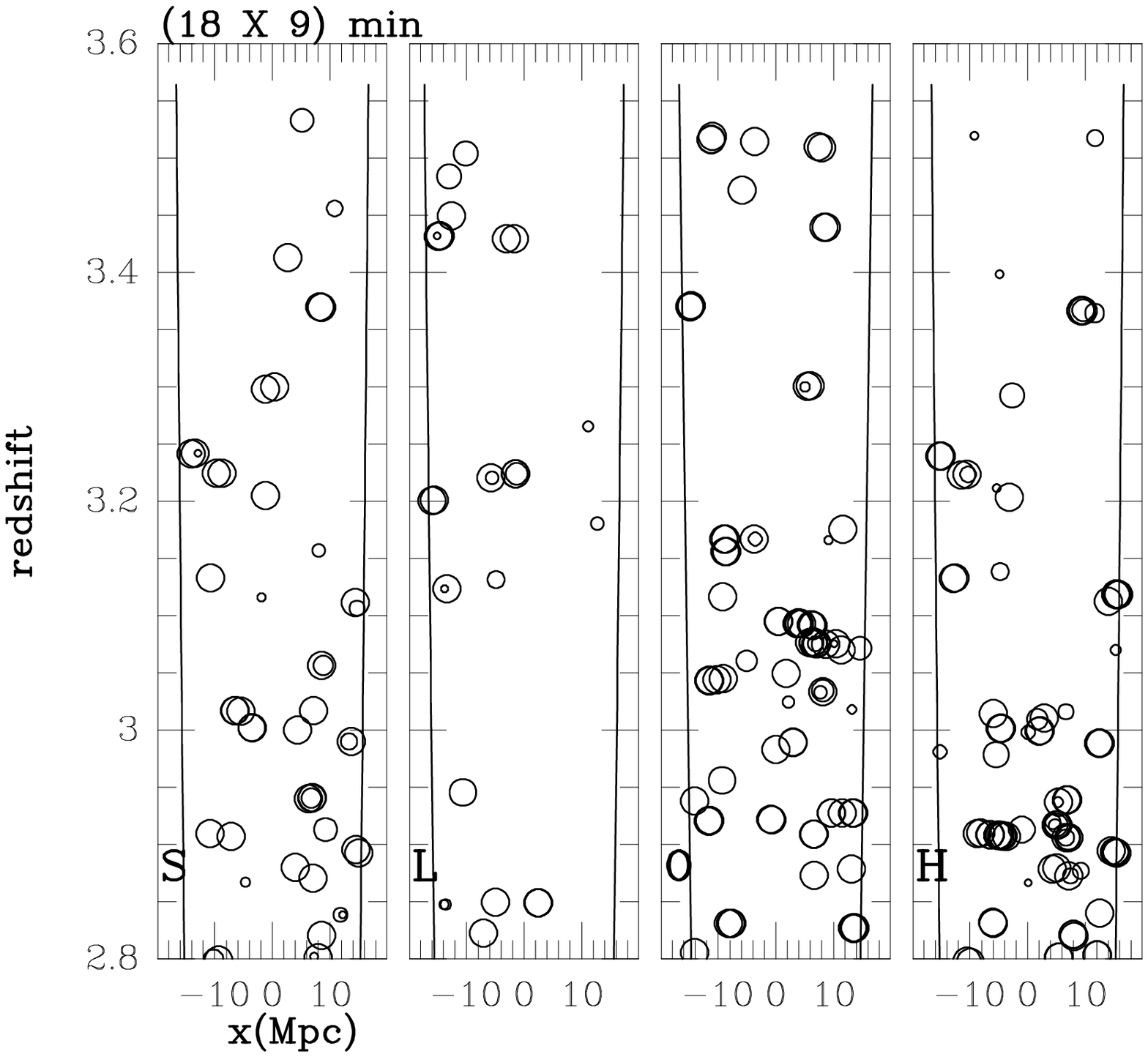}
\hspace{0.1in}
\epsfysize=3.3in\epsfxsize=3.3in\epsfbox{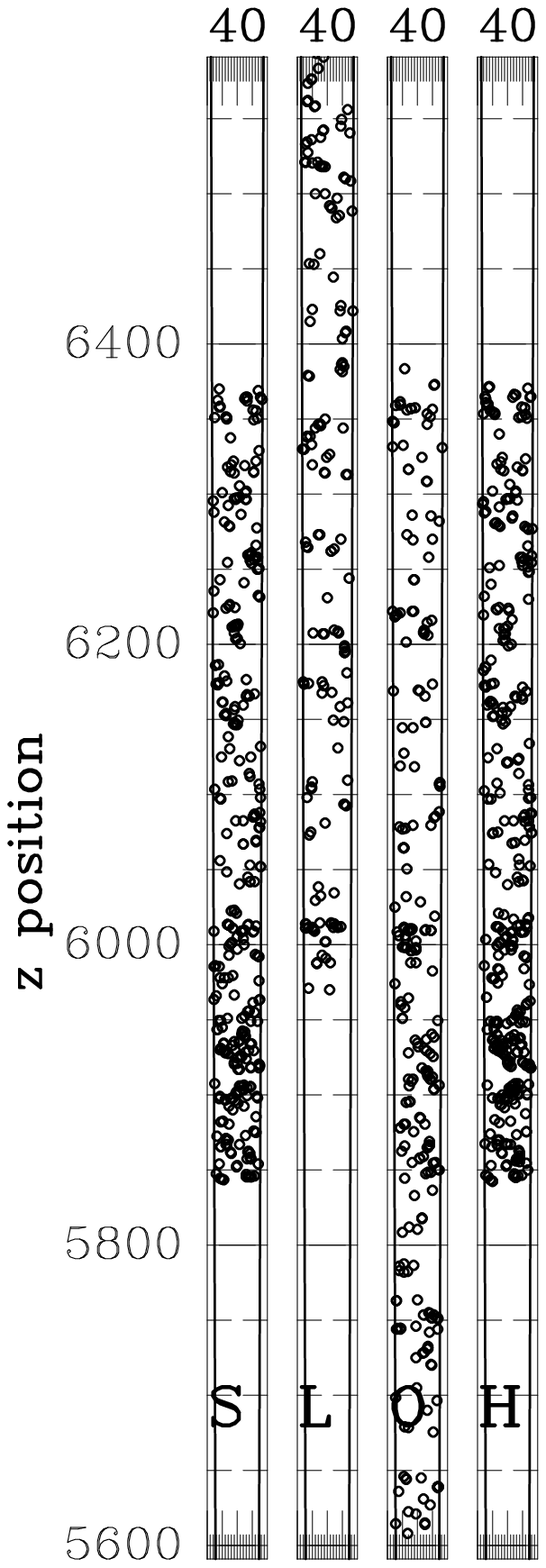}
}
\vspace{-0.95in}
\centerline{
\hspace{1.4in}
\epsfysize=3.5in\epsfxsize=3.5in\epsfbox{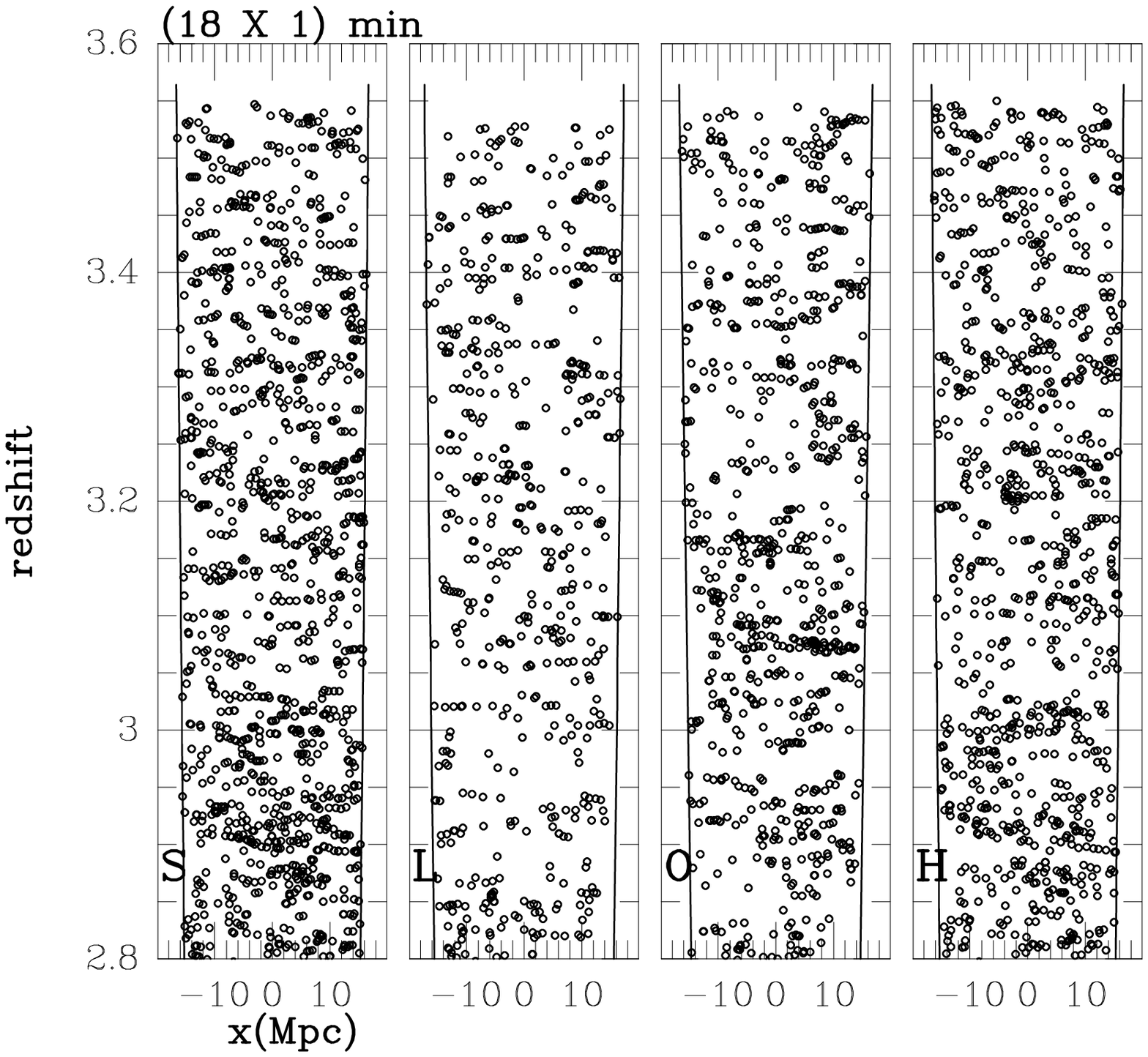}
\hspace{0.1in}
\epsfysize=3.3in\epsfxsize=3.3in\epsfbox{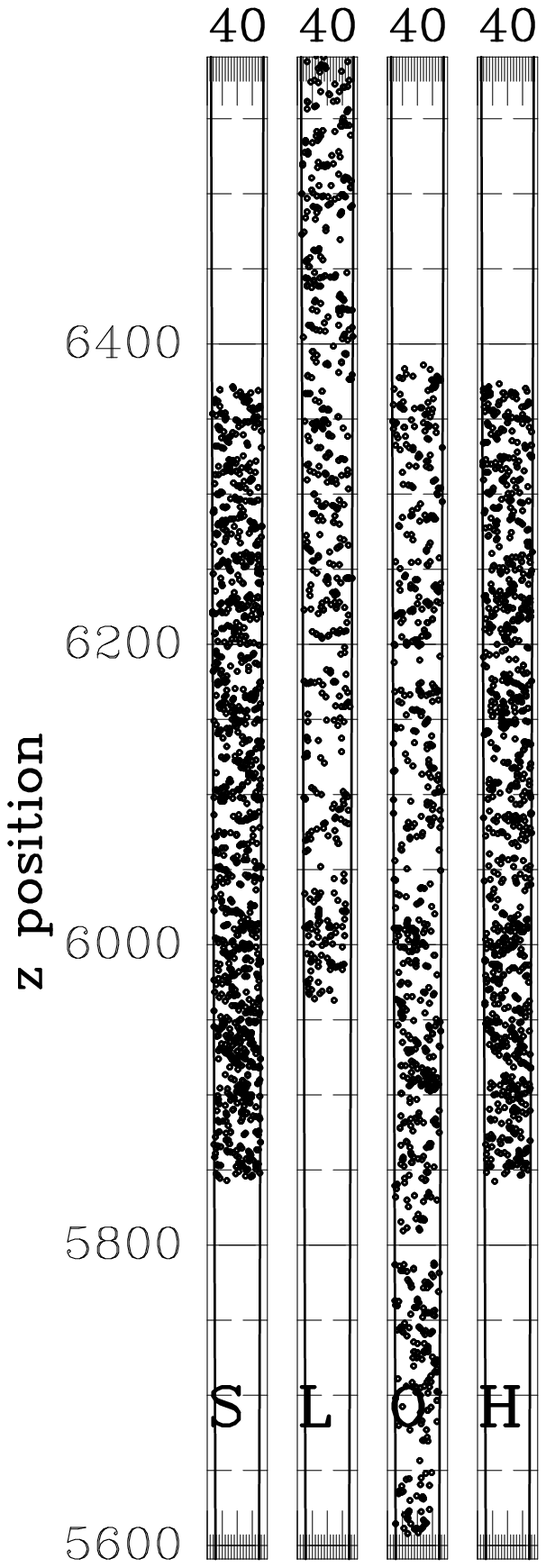}
}
\vspace{-0.25in}
\caption{\small {The redshift-space (left) and comoving-space (right)
distributions of galactic peak-patches. Top left is $18^\prime$ wide
and $9^\prime$ thick, with a $v_{BE}> 290 \kms$ cut, roughly
corresponding to Steidel's damped Ly$\alpha$
systems~{\cite{steidel97}}. The 4 cosmologies shown,
$\{S,\Lambda,O,H\}$CDM, all reveal large scale features, though more
pronounced in OCDM and $\Lambda$CDM.  Top right shows the spatial
distribution, this time for bright galaxies with a $200 \kms$ and
$3\times 10^{10} \msun$ baryonic mass cut, for a region now
$4.5^\prime$ thick. Bottom left and right show a $90 \kms$ cut, for a
$1^\prime$ thick slice. The filaments bridging these dwarf galaxies
define the lower $N_{HI}$ Ly$\alpha$ forest, while halos with $v_{BE}
\sim 30 \kms$ dominate the $N_{HI} \sim 10^{15}$ regime. The forest
hydro simulations cover only $1/8$ of the $40 \mpc$ at high
resolution. This figure therefore emphasizes large scale waves must be
included in our small scale ``shear-patch'' simulations, whose size is
governed by our need to resolve 1~kpc structure in galaxies at
$z$=3. The patches then typically include many $30 \kms$ halos but
only a handful of $90 \kms$ ones.  The currently largest periodic 
simulations with this resolution are not much bigger, but miss the
long waves. We can use these peak-patch/cosmic web simulations to
compare predictions for different cosmologies with the large scale
structure probed by multiple quasar line-of-sight data and long range
velocity space correlations in quasar spectra as well as emerging 
high-$z$ catalogues~{\cite{bwprep}}. }}
\label{ppp} 
\end{figure}

In Fig.~\ref{ppp}, we show the (impressively grand) large scale
clustering in galaxies at various velocity dispersion cuts, in
comoving space and in redshift space. Motivated by \cite{steidel97},
we tiled regions from $z$=2.8 to 3.5 encompassing $18^\prime \times
18^\prime $ with 128$^3$, 40 Mpc boxes. The size was chosen to resolve
``dwarflet'' peak-patches with binding energy $v_{BE} \sim 
30 \kms$. Depending upon cosmology (Table~\ref{table}), 15 to 22 boxes
were needed. (Tiling for open (hyperbolic) universes is an interesting
exercise.) Optimal wavenumber sampling was used
(\S~\ref{sec:methods}), with phase coherent ultralong and short waves
consistently joining box to box.

\begin{table}[h]
\vspace{-0.2in}
\caption{\small {The cosmologies studied are normalized to fit the COBE data,
except for SCDM. The $\{S,\Lambda,O,H\}$CDM hydro runs (first
parameters given) had Gaussian-smoothed $\sigma_\rho (0.5 \, {\rm
Mpc})$=1.05 at $z$=3, and $\Omega_B{\rm h}^2$=0.0125.}}
\vspace{-0.2in}
\begin{center}\small
\begin{tabular}{lllll} 
\hline 
Cosmology &  $\sigma_8$ & H$_0$, Age & $\Omega_{nr}$ & Notes\\
\hline 
SCDM & 0.67,0.44,1.0 & 50, 13 Gyr & 1.0 \\ 
$\Lambda$CDM & 0.93, 0.91 & 70, 13 Gyr & 0.335 & $\Omega_\Lambda$=0.665,$n_s=0.94$, $n_s=1$  \\
OCDM & 0.81, 0.91 & 70, 11 Gyr & 0.37 & $n_s=1$ \\
HCDM & 0.81 & 50, 13 Gyr & 0.8 & $\Omega_{m\nu}=0.2$, 2 degenerate species \\
\hline 
\end{tabular}
\end{center}
\vspace{-0.3in}
\label{table}
\end{table}

Our ``shear-patch'' gas simulations sample the Fig.~\ref{ppp} space only over
small high resolution cores 5 Mpc across (encompassing a fixed
cosmology-independent baryonic mass of $M_b=2\times
10^{11}{\rm\,M_\odot}$), but our images of $N_{HI}, T_{gas}$ \etc
(see~\cite{bwKing97,bwprep}) reveal the cosmic web: In the most
probable shear-patches, the typical IGM at $z\sim 3$ is dominated by
filamentary structures fixed in comoving space with large dwarf
galaxies ($M\sim 10^{10.5} M_{\odot}$, $v_{BE}\sim 90 \kms$,
$N_{HI}\sim 10^{17-19} {\rm cm}^{-2}$) forming at the junctures and
smaller dwarfs ($M\sim 10^{9.2} M_{\odot}$, $v_{BE}\sim 30
\kms$, $N_{HI}$ $\sim 10^{15-17} {\rm cm}^{-2}$) littering the limbs. In
less probable protogalactic shear-patches, the filaments merge, as
part of larger scale filaments.  In protovoid shear-patches, the
webbing is less pronounced. $\Lambda$CDM and OCDM have fatter
ribbon-like filaments with fewer dwarflets compared to SCDM.

\begin{figure}
\vspace{-1.2in}
\centerline{
\epsfxsize=4.5in\epsfysize=3.0in\epsfbox{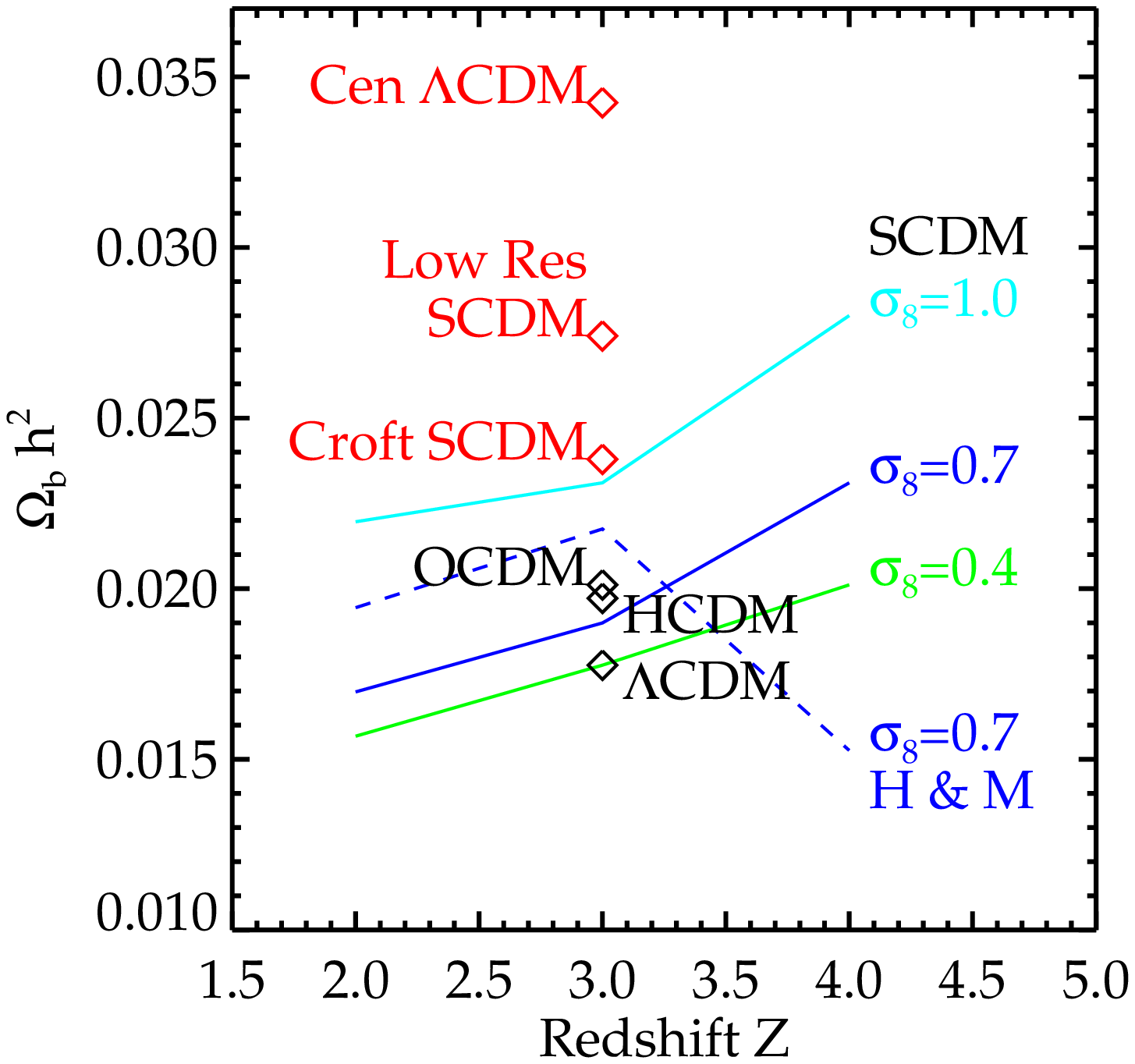}}
\vspace{-0.2in}
\caption{\small $\Omega_b\,{\rm h}^2$ derived from UV flux rescaling
for the 4 cosmological models at $z$=3, and as a function of redshift
for SCDM for the 3 $\sigma_\rho (0.5 \mpc )$ values we tested. A fixed
$J_{-21}$=0.5, typical of $z$=3 estimates, is used, except for the
evolving H\&M (SCDM) curve using the Hardt and Madau UV
flux~{\cite{HM96}}. Low Res SCDM is a simulation with a factor of 3
lower resolution than our standard, which gives incorrect opacity
estimates. ``Cen'' and ``Croft'' refer to simulations discussed in
{\cite{rauch}} and the text. Raising $\sigma_8$ (\ie $\sigma_\rho (0.5
\mpc )$) results in a lower mean opacity by emptying high cross
section filaments into more compact objects. $J_{-21}$ is the flux at
the hydrogen edge in $10^{-21}$ {\it cgs} units.}\label{opacity}
\vspace{-0.05in}
\centerline{
\epsfxsize=4.5in\epsfysize=3.0in\epsfbox{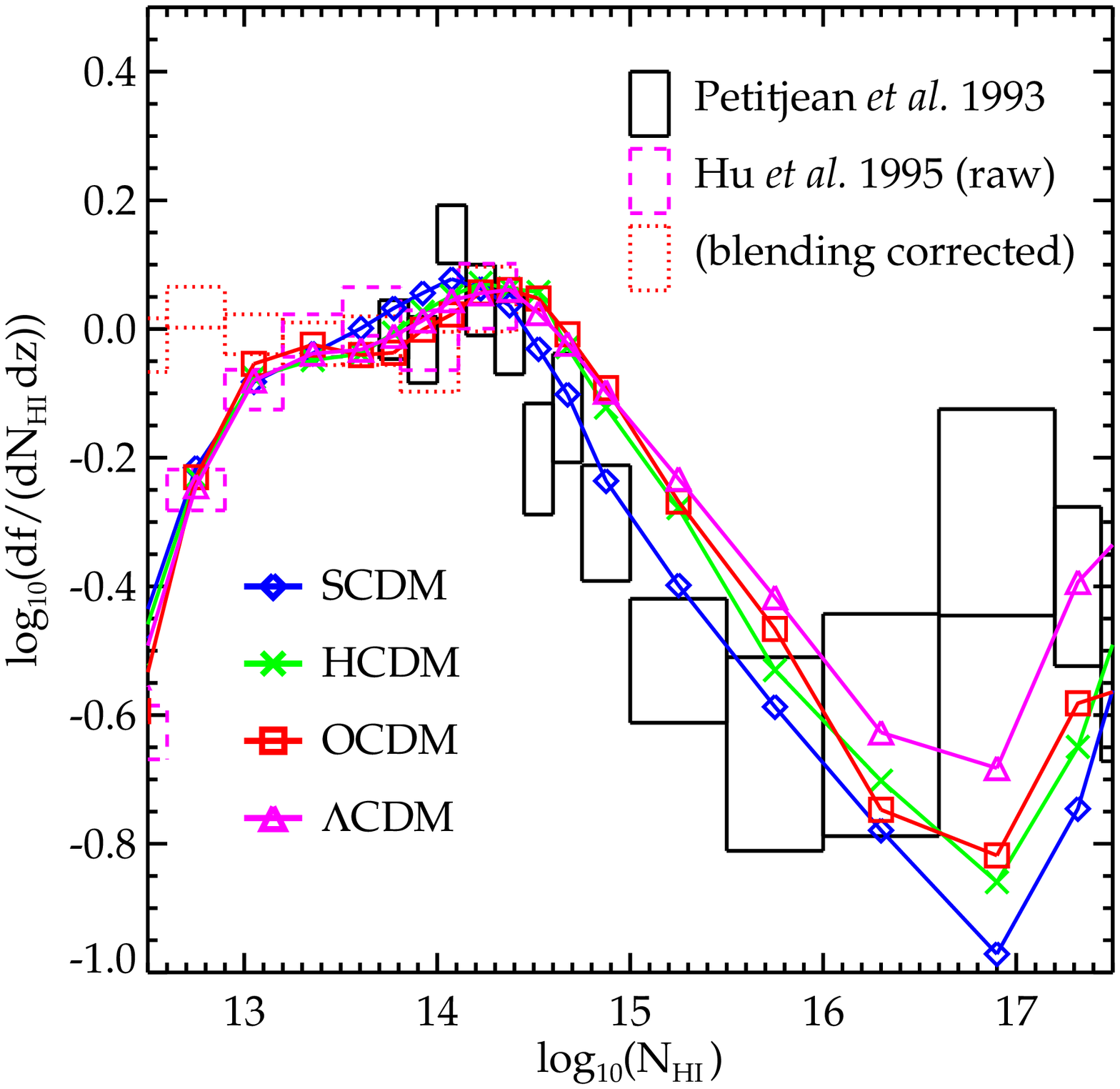}}
\vspace{-0.2in}
\caption{\small Column density distribution at $z$=3, with a $N_{HI}^{-1.46}$
power-law removed, reveals structural details, including the ``break''
at $N_{HI}\sim 10^{14.5} \, {\rm cm}^{-2}$. Results below $10^{16}\,
{\rm cm}^{-2}$ fit the observations well for all 4 cosmologies. This
is partly due to the forgiving nature of opacity rescaling that
prevents differentiation among cosmological models based on this
measure, even when spatial images of the simulations in $N_{HI}$ look
very different. The trend with redshift~{\cite{kim}}  is also
reproduced.}\label{fnhi}
\vspace{-0.2in}
\end{figure}

\section{Ly$\alpha$ Forest Simulations of {``Shear-field Patches''}}\label{sec:methods}

\noindent
{\bf Optimal $k$-space sampling:} The flatness of the power spectrum,
$d\sigma_\rho^2/d\ln k$, for wavelengths $\lambda =2\pi k^{-1} \lta 6
\mpc$ means that small periodic boxes {\it must} exclude substantial
large scale power, and can probe only a restricted set of shearing
patches. We use a hybrid method for applying waves to the initial
conditions, with special attention to the correct inclusion of long
waves~\cite{bm96}: an FFT is used for short waves, but slow direct
transforms are used with power law and log sampling for low $k$, with
switching defined by a $k$-space volume minimization criterion.

\noindent
{\bf Importance Sampling:} Since a single small box is a poor sample
of the universe, we use a number of shearing patch simulations for
each cosmology to be tested, each parameterized by a smoothed
overdensity ($\nu \sigma_\rho (0.5 \mpc)$) and anisotropic shear
tensor. The Gaussian filter $R_{f}$=0.5~Mpc subtends a galaxy mass,
$3\times 10^{11}\Omega_{nr}(2{\rm h})^2 \msun$. We have found that
using a relatively small sample of $\nu$, $\{-1.4, -0.7, 0, +0.7,
+1.4\}$, with the anisotropic shears taken at their mean values given
$\nu$, gives adequate statistical convergence on quantities such as
the frequency distribution of $N_{HI}$, provided the simulations are
given their proper relative weighting. The weighting of $\nu$-bin $i$
is $\propto \int_{\nu_{i-1/2}}^{\nu_{i+1/2}} e^{-\nu^2/2} d\nu$ in
initial condition space, but in the dynamically evolved space it must
be multiplied by the angle on the sky subtended by the region and the 
length in redshift space. The negative $\nu$ (void) cases expand and
thus contribute more, reducing the frequency of lines in the combined
sample relative to what an average density simulation would
give. Overdense regions are more compact and contribute large numbers
of high column lines to the statistics (\eg the $\nu$=1.4
protogalactic patches form collapsed massive galaxies just below $z
\sim 2$).

\noindent{\bf SPH-TreeP$^3$M.} The simulations reported here used a
new code, combining an FFT for the P-M~\cite{couchman} with a
tree-like P-P, which is twice faster than the SPH-P$^3$MG code used in
our earlier work~\cite{bwKing97}, and with more accurate forces. Our
high resolution 5.0 Mpc diameter patches had $50^3/2$ gas and $50^3/2$
dark matter particles (initial grid spacing of 100 comoving kpc),
surrounded by gas and dark particles with 8 times the mass to 8.0 Mpc,
in turn surrounded by ``tidal'' particles with 64 times the mass to
12.8 Mpc. A linearly-evolved self-consistent external shear field was
included to treat ultralong waves. SPH is adaptive: our best
resolution is set by the gravitational softening of 1 kpc.

\noindent{\bf Numerical convergence} is fundamental but has been
largely neglected in previous studies. Simulations with the same
initial conditions, but 100, 50 and 25 comoving kpc initial spacing,
show our standard 100 kpc computations are close to convergence in the
$N_{HI}$ images and in the column density distribution $f(N_{HI})$,
but of course not in small scale dark matter clumping. However, tests
halving the initial resolution to 200 kpc initial spacing show a
significantly lowered mean opacity and fail to produce small dwarf
galaxies (removing objects below $\sim 10^{10} \msun$).

\section{Ly$\alpha$ Forest Response to Varying Physical Parameters} 

We have varied cosmology, UV flux history and power spectrum shape and
amplitude~\cite{bwprep} while keeping the random number choices for
the initial wave phases and the simulation size fixed, to facilitate
comparisons.

\noindent{\bf The flux decrement one-point distribution:} $P(>D_A)$,
$D_A=1-e^{-\tau}$, is determined mainly by the mean opacity and is
insensitive to choices of cosmology. Fitting the observed $P(>D_A)$
well after rescaling is all too easy; we have even shown that
$P(>D_A)$ data from other redshifts can be rescaled to fit
it. (Although we use the full $P(>D_A)$ to rescale, just scaling to
$<D_A>$ gives identical results.)

\noindent{\bf UV flux history:} We have tested how well the much-used
opacity rescaling by $(\Omega_B{\rm h}^2)^2/(J_{-21} H(z))$ actually
works, by evolving identical patches except for the flux history.  We
find it works quite well in the intercloud filamentary regime but can
fail in halos (the traditional clouds), where the {\it Jeans Mass
history} is important, and thus can fail for the higher column
lines.\footnote{We have also shown the {\bf truncated Zel'dovich map}
with a global filter $R_f$ and $\sigma_\rho (R_f) \lta
1$, gives images which differ radically from our
simulations. Allowing $R_f$ to vary reproduces
$f(N_{HI})$ for low $N_{HI}$, a testament to the forgiving nature
of the $f(N_{HI})$ distribution with UV flux rescaling ({\it
cf.}~\cite{hui}).}

\noindent{\bf $\Omega_b$ and UV flux:} Fig.~\ref{opacity} shows the
scaling parameter for the four cosmologies for $J_{-21}=0.5$. At
$z=3$, $(\Omega_b{\rm h}^2)/(2J_{-21})^{1/2}$ ranges from $0.018$ for
$\Lambda$CDM to $0.020$ for OCDM.  Note that our one-third-resolution
results (labeled Low Res SCDM) give a significantly higher estimate;
the estimate labeled Croft \cite{rauch} also had about this
resolution, partly explaining why it is high, but also a colder
thermal history. We also note that the higher flux or lower $\Omega_b$
needed to scale $\Lambda$CDM that we see follows from the slower
expansion rate at high redshift than SCDM has ({\it cf.} the Cen
result~\cite{rauch}).

\noindent{\bf Column density distribution:} Fig.~\ref{fnhi} shows that
with the rescaled UV flux, $f(N_{HI})$ is well-fit by all 4
cosmologies tested. They are visually different, which higher-point
statistics ({\it e.g.}  2-point~\cite{bondzuo}) can better reveal. The
$f(N_{HI})$ were generated from automated Voigt profile fits to
simulated spectra. Each curve represents the weighted composite of the
5 $\nu$ simulations.  Our simulated results do exhibit an evolutionary
trend in the Ly$\alpha$ line counts similar to the Keck observations
of \cite{kim}, a consequence here of the shift in the knee in the
distribution apparent at 10$^{14.5}\, {\rm cm}^{-2}$, which marks the
transition from filamentary IGM material to the halos of dwarf
galaxies.

\noindent{\bf $\sigma_8$ Variation:} For SCDM, we varied the overall
power spectrum amplitude: $\sigma_\rho (0.5 \, {\rm
Mpc},z=3)$=0.7,1.58 as well as 1.05. Although dramatically different
in visual appearance at the same redshift, with UV rescaling the
$f(N_{HI})$ distributions look rather similar at low $N_{HI}$. The
required scaling is larger than for the cosmology variations with
fixed $\sigma_\rho (0.5)$ of Fig.~\ref{opacity}. The median in the
line width, $b$, is $\sim 2 \kms$ higher for 1.58, and $\sim 2 \kms$
lower for 0.7, {\it cf.} the 1.05 case (with median $\approx 30 \kms$, as for
$\{ \Lambda,O,H \}$CDM).

\vfill
\end{document}